\documentclass{emulateapj}

\usepackage{xspace}
\usepackage{graphicx}
\usepackage{rotating}
\usepackage{natbib}
\usepackage{apjfonts}
\def \inte {\textit{INTEGRAL}}
\def \xmm {$XMM$-$Newton$}

\def \sw {$Swift$}

\def \src {IGR\,J16479--4514}

\def \nh {N${\rm _H}$}

\def \hcm {\hbox {\ifmmode $ atom cm$^{-2}\else atom cm$^{-2}$\fi}}

\def \chisq {$\chi ^{2}$}

\def \ATel {Astron.\ Tel.}
\def \apj {ApJ}

\def \apjl {ApJL}
\def \apjs {ApJS}
\def \aap {A\&A}

\def \gcn {GCN Circ.}
\def \mnras {MNRAS}

\newcommand{\be}{\begin{equation}}

\newcommand{\ee}{\end{equation}}

\begin{document}


\title{Dust-scattering halo and giant hard X-ray flare from the  Supergiant Fast X-ray Transient  IGR~J16479$-$4514 investigated with 
 {\itshape XMM-Newton} and {\itshape INTEGRAL}}

\author{V. Sguera\altaffilmark{1}, 
A. Tiengo\altaffilmark{2, 3, 4},
L. Sidoli\altaffilmark{4},
A. J. Bird\altaffilmark{5}
}

\altaffiltext{1}{INAF$-$OAS, Osservatorio di Astrofisica e Scienza dello Spazio, Area della Ricerca del CNR, via Gobetti 101, I-1-40129 Bologna, Italy; vito.sguera@.inaf.it}
\altaffiltext{2}{Scuola Universitaria Superiore IUSS Pavia, piazza della Vittoria 15, 27100 Pavia, Italy }
\altaffiltext{3}{Istituto Nazionale di Fisica Nucleare (INFN), Sezione di Pavia, via A. Bassi 6, 27100 Pavia, Italy}
\altaffiltext{4}{INAF-IASF, Istituto di Astrofisica Spaziale e Fisica Cosmica, Via A. Corti 12, I-20133 Milano, Italy}
\altaffiltext{5}{School of Physics and Astronomy, University of Southampton, University Road, Southampton, SO17 1BJ, UK}

\begin{abstract}
We report results from the analysis of {\itshape XMM-Newton} and {\itshape INTEGRAL}  data of IGR~J16479$-$4514. The unpublished  {\itshape XMM-Newton} observation, performed in 2012, occurred during the source eclipse. No point-like X-ray emission was detected from the source, conversely extended X-ray emission was clearly detected up to a size distance compatible with a dust scattering halo produced by the source X-ray emission before being eclipsed by its companion donor star.  The diffuse emission of the dust-scattering halo could be observed without any contamination from the central point X-ray source, compared to a previous {\itshape XMM-Newton}  observation published  in 2008. Our comprehensive analysis of the 2012 unpublished spectrum of the diffuse emission as well as of the 2008 re-analysed spectra extracted from three adjacent time intervals and different extraction regions (optimized for point-like and extended emission) allowed us to 
clearly disentangle the scattering halo spectrum from the residual point-like emission during the 2008 eclipse. Moreover, the point-like emission detected in 2008 could be separated into two components attributed to the direct emission from the source and to scattering in the stellar wind, respectively.  
From archival unpublished {\itshape INTEGRAL}  data, we identified a very strong ($\sim$3$\times$10$^{-8}$ erg cm$^{-2}$ s$^{-1}$) and fast ($\sim$25 minutes duration)  flare which was classified as giant hard X-ray flare since the measured  peak-luminosity is $\sim$7$\times$10$^{37}$ erg  s$^{-1}$. Giant X-ray flares from SFXTs are very rare, to date only  one has been reported from a different source. We propose a physical scenario to explain the origin in the case of IGR~J16479$-$4514. 

\end{abstract}



        \section{Introduction\label{intro}}

One of the major outcomes  of the  {\itshape INTEGRAL} mission, launched in 2002, has been the discovery of a new class of Supergiant High-Mass X-ray Binaries (SGXBs)  during systematic scans of the Galactic plane: the Supergiant Fast X-ray Transients (SFXTs, Sguera et al. 2005, 2006). They usually host a neutron star orbiting around an early type supergiant star  (Negueruela et al. 2006).  In the X-ray band, they show a rather well defined set of peculiar characteristics  (see Sidoli 2017 for a recent review) which were never  seen from previously known classical SGXBs:  i)  bright ($\sim$10$^{36}$ erg  s$^{-1}$) and fast (few hours to few days) X-ray transient behaviour  ii)  high dynamic ranges of 10$^3$--10$^5$ iii) low duty cycles of  (0.1--5)$\%$ when observed above 
20 keV.

IGR J16479$-$4514 is one of the very first discovered SFXTs. It was newly discovered by   {\itshape INTEGRAL}  in 2003  as an unidentified X-ray transient with no constrained duration (Molkov et  al. 2003), and further  in-depth studies unveiled  its peculiar fast X-ray transient nature (Sguera et al. 2005, 2006). Subsequent near infrared spectroscopy allowed the  identification of its optical counterpart  as a  supergiant star with poorly constrained distance (Chaty  et  al.  2008, Nespoli et al. 2008).  Coley et al. (2015) have recently reported  the best constrained  spectral type (O7 and earlier)  and distance (in the range 4.4--4.6 kpc) for  the companion donor. The   X-ray behaviour  of the source has been investigated with different satellites, e.g.   {\itshape INTEGRAL}  (Sguera et al. 2008),  {\itshape Swift}  (Romano et al. 2008, Bozzo et al. 2009), {\itshape Suzaku} (Sidoli et al. 2013). IGR~J16479$-$4514 is the SFXT with the shortest known orbital period to date ($\sim$3.3 days, Jain et al. 2009), interestingly  a superorbital period of $\sim$11.88  days has been detected as well from {\itshape Swift} and {\itshape INTEGRAL} observations (Corbet et al. 2013, Drave et al. 2013). The source duty cycle value is one of  the highest among the entire sample of firm SFXT ($\sim$3.3$\%$, Sidoli $\&$ Paizis 2018).  IGR~J16479$-$4514 is known to display X-ray eclipses  as discovered during   an {\itshape XMM-Newton} observation  which covered part of the eclipse ingress along with part of its total phase (Bozzo et a. 2008). 

Here we report new results on IGR J16479$-$4514  as obtained from an  unpublished targeted  {\itshape XMM-Newton} observation and from 
archival  {\itshape INTEGRAL} data.

\section{Data analysis}

Throughout the paper, spectral analysis was performed using
XSPEC version 12.9.0, with the photoelectric absorption model based on the Balucinska-Church \& McCammon (1992) cross-sections and Anders \& Grevesse (1989) solar abundances. To properly use the \chisq\ statistics in model fitting, all the spectra were grouped to a minimum of 30 counts per energy bin. Unless stated otherwise, errors are quoted at the 90 per cent confidence
level for one single parameter of interest.

\subsection{INTEGRAL}

For our study, we used data collected with the ISGRI detector (Lebrun et al. 2003) which is  the lower energy 
layer of the IBIS coded mask telescope (Ubertini et al. 2003) onboard {\itshape INTEGRAL} (Winkler et al. 2003).

The IBIS/ISGRI public data archive (from revolution 30 to 1500, i.e. from approximately January 2003  to January 2015)  
has been specifically searched for very powerful hard X-ray flares from IGRJ16479$-$4514
detected at Science Window level (ScWs, $\sim$ 2,000 seconds duration).  In particular, the data set consists of
9,578 ScWs where IGR~J16479$-$4514  was  within the total instrument 
field of view (FoV) of 29$^{\circ}\times29^{\circ}$ (down to zero response), i.e. regardless of its  off-axis angle. 
For the sake of completeness, we note that a 12$^{\circ}$ limit  is generally applied because the
response of IBIS/ISGRI is not well modelled at large off-axis values and this  may introduce a systematic error in
the measurement of the source fluxes.  However our specific aim is to  search for exceptionally powerful outbursts (i.e. flux greater than at least 500 mCrab as explained later in the text) which are, likely, very rare events. To increase the probability of finding such events, we deliberately  did not place any requirement  on the off-axis angle of the source: the larger the considered FoV, the greater the chance of serendipitously detecting
a powerful random flare. 
Because the considered flares are  exceptionally bright, their   eventual detection and flux would  be confidently determined even when their position is in the partially coded region of the FoV.  IBIS/ISGRI flux maps for each ScW were generated in the 18--60  keV band  using the offline scientific analysis  software  OSA 10.2. Count rates at  the position  of the source were extracted from individual ScW flux maps.  This approach is particularly efficient in unveiling  fast transient flares  lasting only a very few hours, since the search occurs on the same timescale as the outbursts themselves. In order to specifically search for the most powerful flares,  we adopted a conservative source count rate threshold of  80 cts/s (18--60 keV) 
as measured in the single ScW containing the peak of the flare; it translates into a 18--60 keV flux of  $\sim$ 500 mCrab   by assuming a Crab-like energy
spectrum with photon index equal to 2.1. Throughout the paper, the conversion from the source count rate to mCrab flux  has been obtained by using the most recent Crab observations at the time of writing (revolution 1856, August 2017), i.e. 1 Crab=158.13 cts/s (18--60 keV). The search was initially performed in the energy band 18--60 keV,  then when a candidate fast flare was found  we have  also checked the detection in different energy ranges  (i.e. 17--30 keV, 20--40 keV) in order to maximize the source  best significance detection.  After finding a flare, we performed a more detailed timing and spectral analysis. For the latter, we used the standard 13 energy channel response matrix available at the {\itshape INTEGRAL} Science Data Centre (ISDC). 

In addition to the investigation of the entire public archive, we have also specifically analyzed an archival targeted {\itshape INTEGRAL } observation of  IGR~J16479$-$4514 (revolution 1203, August 2012), not yet published in other works, which was simultaneous to an unpublished  {\itshape XMM-Newton} observation whose results are reported for the first time in section 3.1.

\subsection{XMM-Newton}

\begin{figure}
\begin{center}
\centerline{\includegraphics[width=6cm]{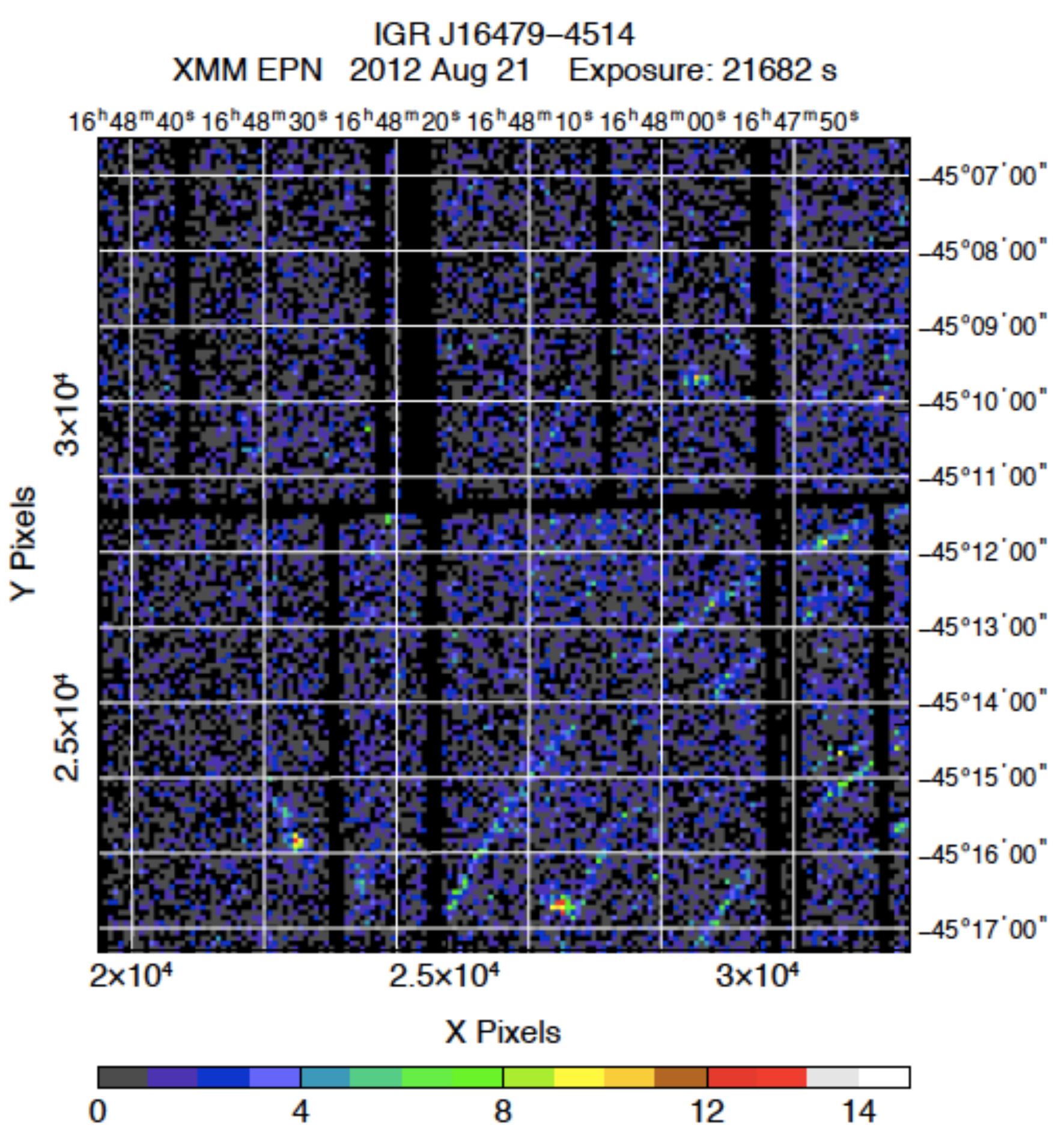}}
\centerline{\includegraphics[width=6cm]{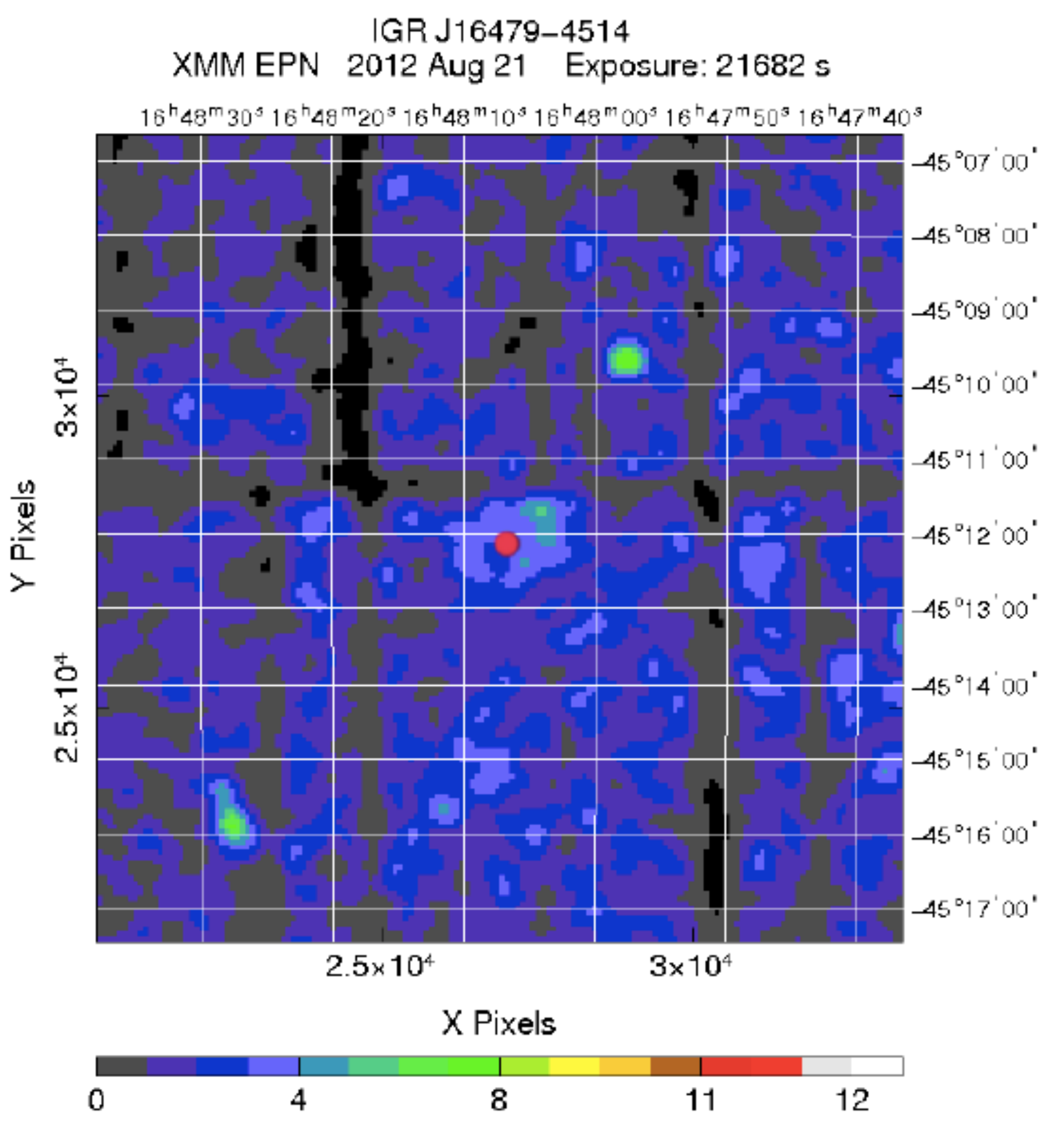}}
\caption{\src\ sky position observed by \xmm/EPIC pn, in the energy ranges 0.5-10 keV 
(top panel, close-up view around the target position) 
and 1-3 keV (bottom panel, where the image was smoothed with a Gaussian with a width $\sigma$=2 pixels). 
The small red circle  marks the \src\ infrared position (2MASS).}
\label{fig:xmm_ima}
\end{center}
\end{figure}

\begin{figure}
\begin{center}
\centerline{\includegraphics[width=3cm,angle=-90]{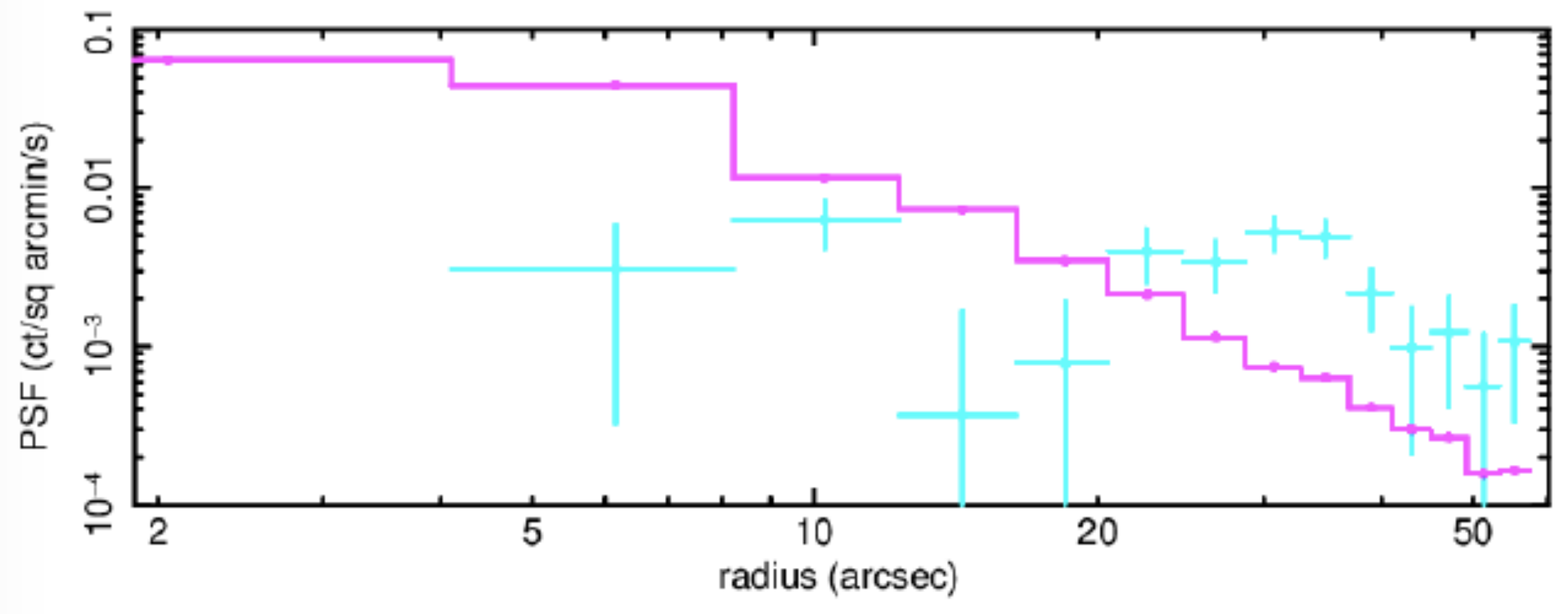}} 
\caption{Surface brightness profiles of the X--ray emission within 1$'$ from the source 2MASS position, 
in the energy range 1--3 keV, 
compared
with the EPIC pn point spread function in the same energy range (solid line).}
\label{fig:xmm_psf}
\end{center}
\end{figure}

The {\itshape XMM-Newton} $Observatory$ (Jansen et al. 2001) carries three 1500~cm$^2$ X--ray
telescopes, each with an European Photon Imaging Camera (EPIC; 0.2--12 keV)
at the focus. One of the EPIC  uses a pn CCD (Struder et al. 2001)
while the other two are equipped with MOS CCDs (Turner et al. 2001).

We discuss here two observations, one performed in 2012 (never reported in the literature), and another one done in 2008 (Bozzo et al. 2008), 
which we have re-analysed here for comparison to obtain a more complete view of the phenomenology during the X--ray eclipse.

IGR\,J16479--4514 was observed with {\itshape XMM-Newton} on 2012 August 21 (from 11:31:03 to 18:04:47, UTC) for a net exposure of $\sim$23.9~ks 
during an observation   simultaneous to part of the above mentioned {\itshape INTEGRAL} observation. The particle background during the {\itshape XMM-Newton} observation was very stable.
All the three EPIC cameras operated in 
Large Window mode and
adopted the medium filter.
Data were reprocessed using version 16.0.0 of the Science Analysis Software (SAS) 
with standard procedures and the appropriate calibration files. 
The response and ancillary matrices were generated using the SAS metatask {\em rmfgen} and {\em arfgen}.
Event patterns 0-4 were used when extracting EPIC pn products, and patterns 0-12 for both the MOS cameras. 

The  {\itshape XMM-Newton} exposure was affected by  stray light contamination produced by a bright source outside the field of view, as  is clearly evident in Fig.~\ref{fig:xmm_ima} (top panel).  The contaminating source is the LMXB  GX~340$+$0, which is $\sim$35 arcmin from IGR\,J16479--4514. Since 
GX~340$+$0 is highly absorbed, its contribution at low energies is negligible (Fig.~\ref{fig:xmm_ima}, bottom panel). 

The source was observed also in 2008, from March 21 14:40 to March 22 01:30, for a net exposure time of about 30~ks, with EPIC pn in Small Window mode and the MOS in Full Frame. The MOS observations suffered from stray light contamination also during this observation.

\section{Results}

\subsection{XMM-Newton}

The 2012 {\itshape XMM-Newton} observation was also performed during the source eclipse, according to the most recent ephemeris reported by 
Coley et al. (2015): orbital period P$_{orb}$=3.31961~days, mid-eclipse time = 55081.571 MJD, eclipse duration, $\Delta\phi$=0.177, 
eclipse at orbital phase $\phi$=0.0.

The EPIC pn exposure covers the time interval 56160.4995--56160.75047 (MJD), implying an orbital phase range $\Delta\phi$=0.02--0.09. 
The eclipse ingress started on  56160.15~MJD, 0.35~days before the start of the EPIC pn exposure.
At odds with the {\itshape XMM-Newton} observation performed in 2008, which caught 
 the source during the eclipse ingress (Bozzo et al. 2008),
\src\ was not detected by {\itshape XMM-Newton}  (Fig.~\ref{fig:xmm_ima}):
running the detection SAS metatask {\em edetect\_chain} on the three EPIC cameras separately,
the three final source lists did not include any point source positionally coincident with \src.

Nevertheless, there is evidence for the presence of diffuse X--ray emission, mainly at low energies, around the source position in the EPIC pn exposure 
(Fig.~\ref{fig:xmm_ima}; net exposure of 21.7~ks). 
%
This finding is confirmed by the radial distribution of the X--ray emission extracted
from within  1~arcmin from the 2MASS source position. We compared this profile with 
the distribution expected from a point-like source (Fig.~\ref{fig:xmm_psf}), resulting in a clearly diffuse emission. 
Since soft X-rays are efficiently scattered by interstellar dust, this diffuse emission might be the X-ray halo produced by the bright X-ray emission of IGR~J16479$-$4514 before eclipse scattered by a large number of dust grains along the line of sight. Depending on the scattering angle and the distance of the dust, due to the longer path, the scattered emission can reach the observer with a time delay up to several days (see, e.g., Pintore et al. 2017 and references therein). 

\begin{figure}
\includegraphics[height=8cm, angle=0]{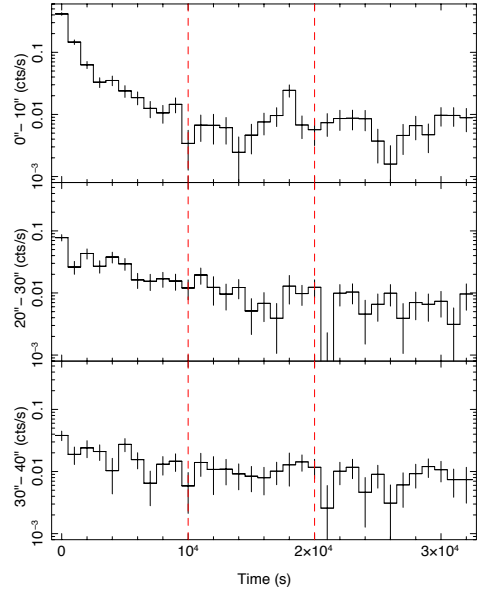}
\caption{EPIC pn  background-subtracted light curve of the 2008 observation extracted in the 1.5--5 keV energy band from a circle of 10 arcsec radius (top panel) and two annuli of inner and outer radii of 20 and 30 arcsec (middle panel) and 30 and 40 arcsec (bottom panel), centered at the \src\ position. The bin time is 1000 s and the vertical lines indicate the three time intervals used for spectral analysis (interval 1, 2, 3 as reported in Table 1).
}
\label{fig:lc08}
\end{figure}



\subsubsection{XMM-Newton spectroscopy}

\begin{table*}
        \centering
        \caption{Best-fit parameters of {\itshape XMM-Newton} spectra with the model reported in equation 1 ($\chi^2$=164.7 for 176 degrees of freedom). Interval 1, 2 and 3 correspond to the  pn spectra of the time intervals of the 2008 observation, as defined in Fig.~\ref{fig:lc08}, and Observation 2 to the  MOS and pn spectra of the full 2012 observation. }
        \label{tab:specpar}
        
        \begin{tabular}{ccccccccc}
        \hline
         & &  \multicolumn{3}{c}{POINT SOURCE (10 arcsec) } &          \multicolumn{4}{c}{DIFFUSE EMISSION (20--40 arcsec)}\\
         & & Interval 1 & Interval 2 & Interval 3 & Interval 1 & Interval 2 & Interval 3 & Observation 2 \\
        \hline
        $N_{\rm H}$  & & 6.6$^{+2.5(a)}_{-2.4}$ & 6.6$^{(a)}$ & 6.6$^{(a)}$ & 6.6$^{(a)}$ & 6.6$^{(a)}$ & 6.6$^{(a)}$ & 6.6$^{(a)}$ \\
        \multicolumn{2}{l}{(10$^{22}$ cm$^{-2}$)} &  & & & & & & \\
        \hline
        $N_{\rm H1}$  & & 33$\pm$5$^{(a)}$ & -- & -- & 33$^{(a)}$ & -- & -- & -- \\ 
        \multicolumn{2}{l}{(10$^{22}$ cm$^{-2}$)} & &  & & & & & \\
        \hline
        $\alpha$ & &  1.2$^{+0.2(a)}_{-0.1}$ & 1.2$^{(a)}$ & 1.2$^{(a)}$ & 1.2$^{(a)}$ & 1.2$^{(a)}$ & 1.2$^{(a)}$ & --\\ 
         & & & &  & & & & \\
        \hline
        $I_{\rm 1}^{(b)}$ &  & 39$^{+7}_{-6}$ & -- & -- & 3.8$^{+1.3}_{-1.2}$ & -- & -- & --\\
        \multicolumn{2}{l}{(10$^{-12}$ erg cm$^{-2}$ s$^{-1}$)} & & & & &  & & \\
        \hline
        $I_{\rm 2}^{(b)}$  & & 2.8$^{+1.5}_{-1.1}$ & 0.65$^{+0.16}_{-0.14}$ & 0.51$\pm$0.12  & 1.2$^{+0.5}_{-0.6}$ & 0.32$^{+0.12}_{-0.14}$  & 0.13$^{+0.10}_{-0.12}$ & -- \\
        \multicolumn{2}{l}{(10$^{-12}$ erg cm$^{-2}$ s$^{-1}$)}  & & & & & & & \\   
        \hline
        $\beta$ &  & -- & -- & -- & 4.8$^{+1.5(a)}_{-1.3}$ & 4.8$^{(a)}$ & 4.8$^{(a)}$ & 4.8$^{(a)}$ \\
         & & & & & & & & \\
        \hline
        $I_{\rm halo}^{(b)}$ & & -- & -- & -- & 0.37$^{+0.50}_{-0.31}$ & 0.32$^{+0.31}_{-0.17}$ & 0.33$^{+0.29}_{-0.16}$ & 0.10$^{+0.08}_{-0.04}$ \\ 
        \multicolumn{2}{l}{(10$^{-12}$ erg cm$^{-2}$ s$^{-1}$)} & &  & & & & & \\
        \hline
        $E_{\rm ln1}$ & &  6.36$^{+0.06(a)}_{-0.04}$  & 6.36$^{(a)}$  & 6.36$^{(a)}$  & -- & -- & -- & -- \\
        \multicolumn{2}{l}{(keV)} & &  & & & & & \\ 
        \hline
        $I_{\rm ln1}^{(c)}$ & &  10$\pm$3$^{(a)}$  & 10$^{(a)}$  & 10$^{(a)}$ & -- & -- & -- & -- \\
        \multicolumn{2}{l}{(10$^{-6}$ photons cm$^{-2}$ s$^{-1}$)} & & & & & &  & \\ 
        \hline
        $E_{\rm ln2}$ & &  6.94$\pm$0.09$^{(a)}$ & 6.94$^{(a)}$ & 6.94$^{(a)}$ & -- & -- & -- & -- \\
        \multicolumn{2}{l}{(keV)} & & & & & &  & \\ 
        \hline
        $I_{\rm ln2}^{(c)}$ & &  8$\pm$3$^{(a)}$ & 8$^{(a)}$ & 8$^{(a)}$ & -- & -- & -- & -- \\
        \multicolumn{2}{l}{(10$^{-6}$ photons cm$^{-2}$ s$^{-1}$)} & & & & & & & \\ 
        \hline
        \hline
    \end{tabular}
\begin{flushleft}    
{$^{(a)}$ \footnotesize Linked parameters}\\
{$^{(b)}$ \footnotesize Unabsorbed flux in the 2--10 keV energy band.}\\
{$^{(c)}$ \footnotesize Total photon flux in the line.}
\end{flushleft}
\end{table*}


\begin{figure}
\includegraphics[height=10cm]{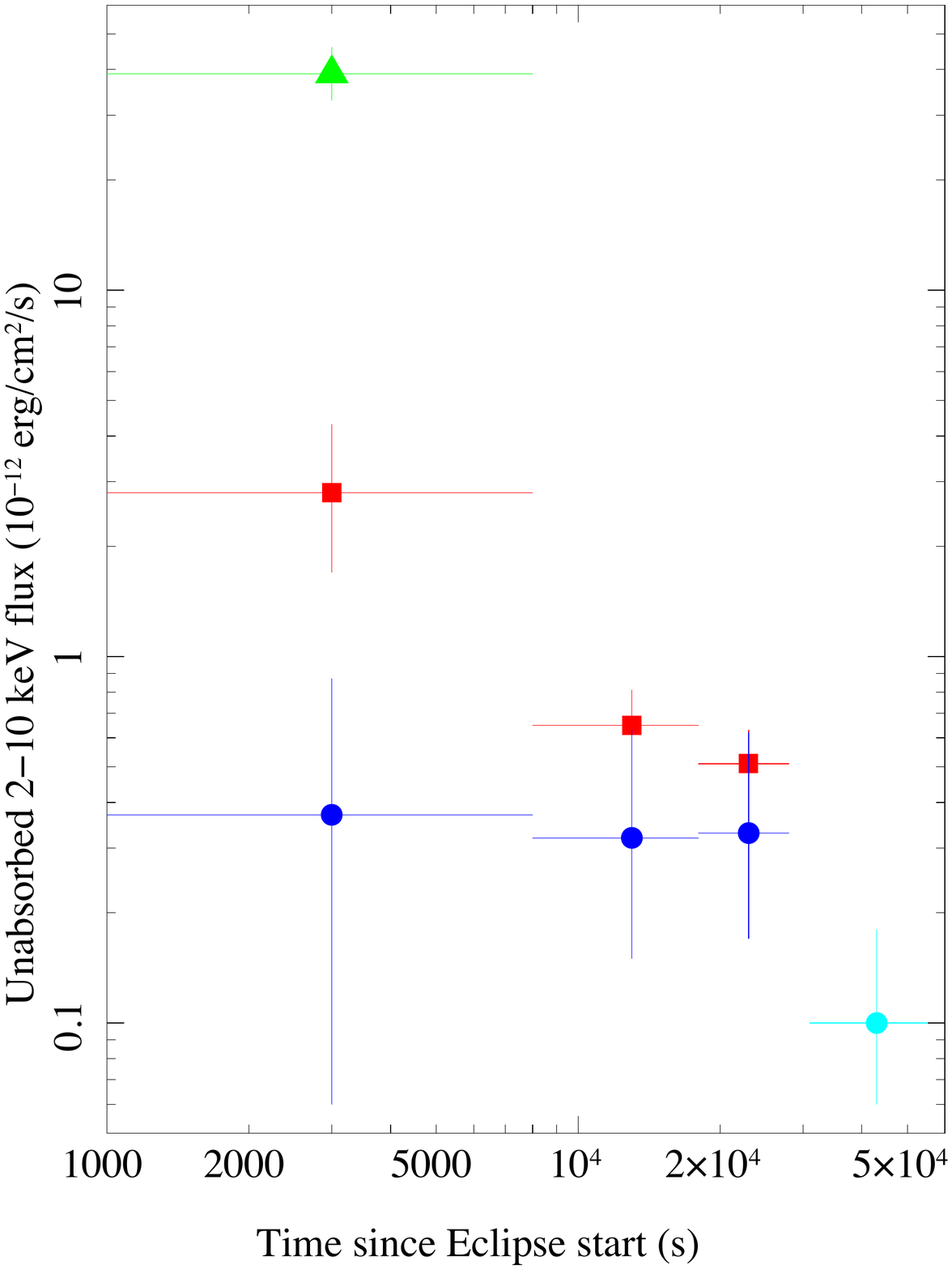}
\caption{Unabsorbed 2--10 keV flux of different components (as reported in Table~\ref{tab:specpar}) in the 2008 (within 30 ks since the eclipse start) and 2012 (more than 30 ks after the eclipse start) 
spectra of the point source (within 10 arcsec; red squares and green triangle) and diffuse emission (from 20 to 40 arcsec; circles). The cyan circle indicates the total flux of the diffuse emission in 2012 and the blue circles are the fluxes of the halo component in the 2008 diffuse emission spectra. The green triangle is the flux of the direct component, which is significantly detected only in the  interval 1 of the 2008 point source emission, whereas the red squares are the fluxes of the point source scattered component detected in the 2008 observation. 
}
\label{fig:fluxes}
\end{figure}

 The EPIC pn, MOS1 and MOS2 spectra of the diffuse X--ray emission observed in 2012 were extracted from an annular region with inner radius of 20 arcsec and outer radius of 40 arcsec, centered at  the \src\ position.  The background for the three cameras was extracted from two rectangular regions located in the same 
CCD and not contaminated by stray light, with a total area of 7,200 arcsec$^2$.  


To test the dust-scattering halo hypothesis and search for spectral evolution during eclipse, we compared the background-subtracted spectrum  to the corresponding spectra extracted in 
different time intervals 
from the 2008  {\itshape XMM-Newton} observation, where the presence of a dust halo was already suggested (Bozzo et al. 2008). 
 In this case, only pn data can be used because the stray light from GX~340$+$0 in the MOS cameras  reached the region where IGR~J16479$-$4514 is located.
That observation started at the beginning of the eclipse and 
Bozzo et al. (2008) performed a time-resolved spectral analysis by dividing the observation into two time intervals, corresponding to the initial 4 ks of high, rapidly decaying flux and the remaining 28 ks of low, almost constant, flux.
To better separate the direct and scattered components, that are expected to have a different spatial distribution and temporal evolution, 
we extracted
and analysed spectra from three time intervals (see Fig.~\ref{fig:lc08}), sampling also the possible evolution during the low flux state, and from two different extraction regions:
a 10 arcsec circle to maximize the contribution from the central point source and the same annulus used for the 2012 spectrum to study the evolution of the possible dust-scattering halo.

Similarly to Bozzo et al. (2008) we adopt a spectral model consisting of three power-law, two Gaussian and two photoelectric absorption components, combined in this way:

\begin{eqnarray}
I(E) & = & e^{\sigma(E) N_{\rm H}}[e^{\sigma(E) N_{\rm H1}}I_{\rm 1} E^{-\alpha} +I_{\rm 2} E^{-\alpha}+ \label{eq:xspec} \\ 
&& I_{\rm halo} E^{-\beta} + I_{\rm ln1} e^{-(E-E_{\rm ln1})^2/(2\sigma_{\rm ln1}^2)} + \nonumber \\ 
&& I_{\rm ln2} e^{-(E-E_{\rm ln2})^2/(2\sigma_{\rm ln2}^2)}] \nonumber
\end{eqnarray}

We use this model to simultaneously fit the nine available spectra (extracted both from a 10 arcsec circle and a 20--40 arcsec annulus in the three time intervals of the 2008 observation, only from a 20--40 arcsec annulus in the 2012 observation), linking together several parameters and forcing to a null normalization all the spectral components not significantly required to model each spectrum. The width of the two emission lines was 
 fixed to the instrumental spectral resolution. The resulting best-fit parameters are reported in Table~\ref{tab:specpar}.

The spectral lines were significantly detected, with consistent parameters, only in the point-source emission. On the other hand, the steepest power-law component, that can be interpreted as the contribution from the dust scattering halo, was required only in the fit of the spectra of the diffuse component. The other two power-law components have the same slope, but one of them is detected only during the first time interval and requires an additional photoelectric absorption component. These two components can be attributed to direct and small-scale scattered emission from the point source, but, due to the relatively coarse PSF of {\itshape XMM-Newton} mirrors, during the 2008 observation they heavily contaminate the diffuse emission extracted from $>$20 arcsec. The normalizations of the three power-law components as a function of time since eclipse start are shown in Fig.~\ref{fig:fluxes}.

\subsection{INTEGRAL}

\subsubsection{Archival search for powerful flares}

Our search found  a very powerful fast hard X-ray flare from IGR~J16479$-$4514. Here we report on the collected  results. 
It was detected  with a significance of  $\sim$ 9$\sigma$ (18--60 keV) during only one ScW (n. 43) in revolution 663, at MJD 54544.96. Notably the source off axis  angle was  $\sim$ 17$^{\circ}$.5. The measured average 18--60 keV flux was 630$\pm$70 mCrab or $\sim$8.2$\times$10$^{-9}$ erg 
cm$^{-2}$ s$^{-1}$. No detection was obtained in the higher energy band 60--100 keV. The ScW spanned a time range from  2008,  March 19 22:37 (UTC) to March 19 23:30 (UTC).  

Figure 5 shows the 18--60 keV IBIS/ISGRI source light curve extracted from the entire ScW with a  bin time of 50 s. A single flare is very evident, being characterized by a duration of  $\sim$ 25 minutes as well as a  remarkable peak  count rate of $\sim$ 330 cts/s which translates into a flux of 2.10$\pm$0.26 Crab  (or $\sim$2.7$\times$10$^{-8}$ erg cm$^{-2}$ s$^{-1}$).   From the light curve we note that the duration of the flare is a bit shorter than the entire duration of the ScW.  Bearing this in mind, we performed an imaging analysis with the
good time interval (GTI) by considering only the time interval of the transient activity (i.e. from $\sim$ 2,700 s to $\sim$ 4,300 s on  x-axis in Fig. 5). By doing so, the source  was significantly detected at 
13.3$\sigma$  level (18--60 keV) despite being located at large off axis angle,  as it can be seen in  the  significance image in Fig. 6.

Unfortunately,  the source was out of the  IBIS/ISGRI FoV in observations performed immediately before and after the   ScW containing the 
detection of the flare, and thus it is  not possible to investigate with IBIS/ISGRI if further flares  took place or not.  To this aim, the BAT hard X-ray transient monitor on board {\itshape Swift}  is particularly suited since it provides a much  more  continuous  coverage thanks to its very large FoV.  We retrieved the 15--50 keV {\itshape Swift}/BAT light curve averaged on orbital time-scale ($\sim$ 96 minutes) from the publicly available BAT monitor web page (Krimm et al. 2013).  As shown in Fig. 7, 
no additional X-ray activity has been detected  immediately before and after the occurrence of the flare.  
The shadowed area indicates the  flaring activity detected by  IBIS/ISGRI  during the single ScW.  We remark that both {\itshape INTEGRAL}/IBIS and {\itshape Swift}/BAT provides  complementary information: i) the continuous  coverage by BAT allows us  to exclude the occurrence of further X-ray activity before and after the flare, ii) the finer  IBIS/ISGRI light curve (50 s bin time), with respect to the BAT light curve (95 minutes bin time),  allows to unambiguously catch and estimate its exact peak-time and  peak-flux, respectively. 

\begin{figure}
\begin{center}
\includegraphics[height=7cm]{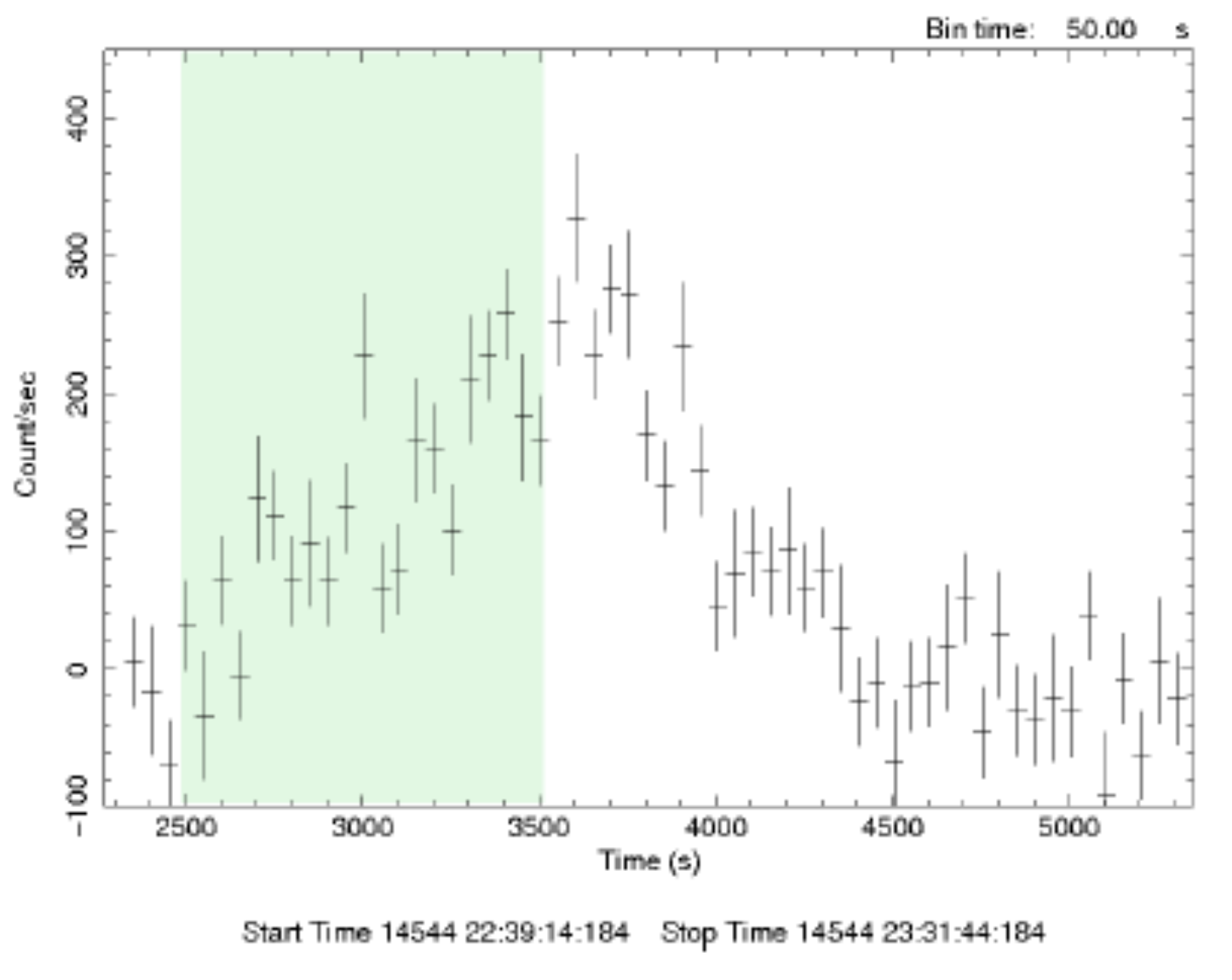}
\caption{IBIS/ISGRI light curve (18--60  keV, 50 s bin time) of the flaring activity from IGR~J16479$-$4514 as extracted from the single  ScW n. 43 in revolution 663. The shadowed green area indicates the portion of the flare  covered by the \sw~observation reported by Romano et al. (2008)}
\end{center}
\begin{center}
\includegraphics[height=7.5cm,angle=0]{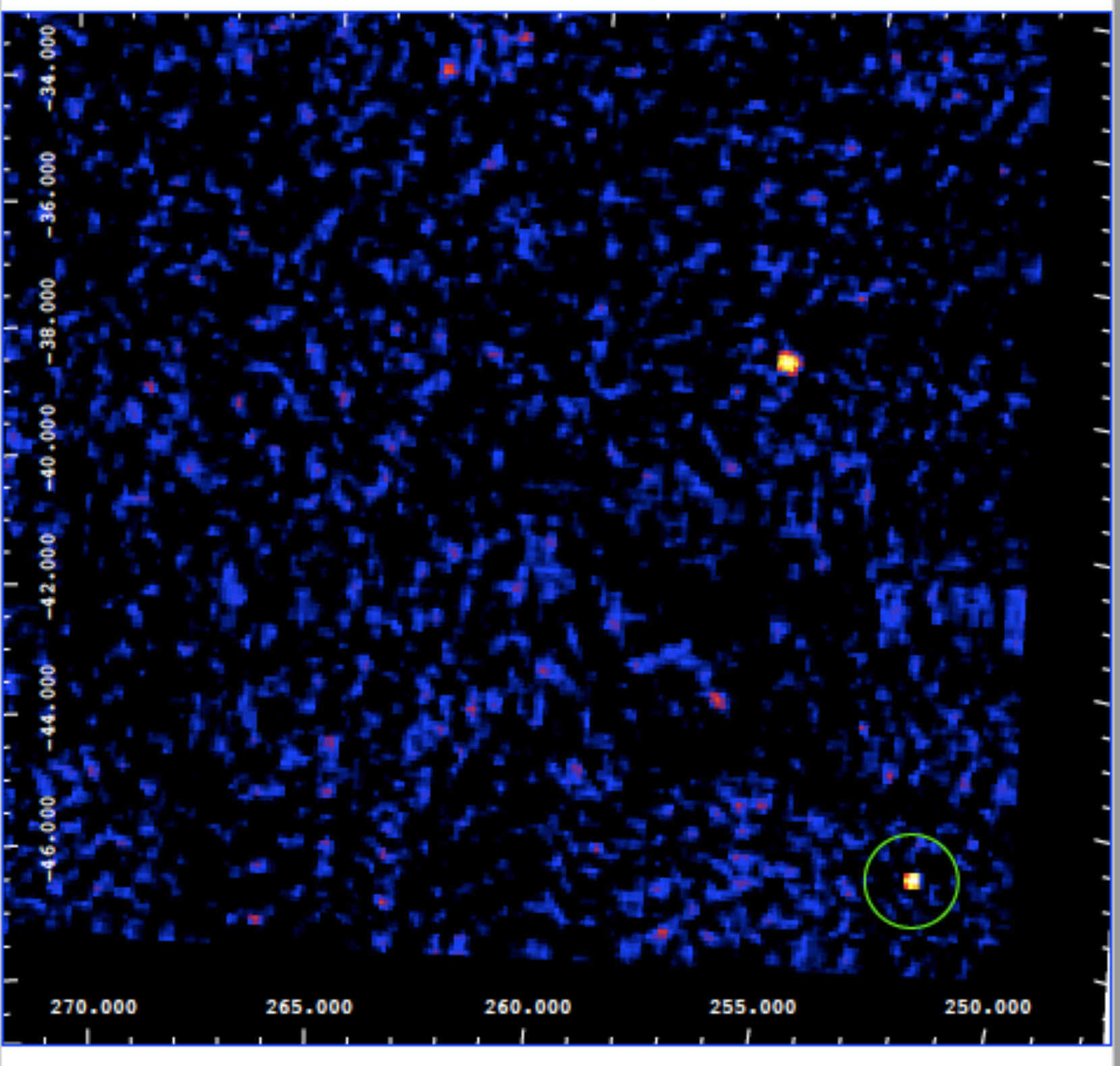}
\caption{IBIS/ISGRI significance map (18--60 keV) of the single ScW  n.43 in revolution 663. 
IGR~J16479$-$4514 (encircled) is  clearly detected at 13.3$\sigma$  level towards the edge of the FoV.
The other bright source detected in the field is  the HMXB  4U~1700$-$377.}
\label{fig6}
\end{center}
\end{figure}

Fig. 8 shows the IBIS/ISGRI light curve of the source (18--60 keV) phase-folded on the orbital period. 
It is clearly modulated by the deep X-ray eclipse (yellow shadowed area in Fig. 8) which is consistent with the compact object  being totally
eclipsed by the supergiant companion. 
According to the most recent and precise ephemerides of Coley et al. (2015), i.e. epoch of the  mid-eclipse MJD 55081.57 as $\phi$=0, 
then the powerful X-ray flare detected by  {\itshape INTEGRAL}  took place at orbital phase $\phi$$\sim$0.65 (grey thick line in Fig. 8).  
 
 We searched  for possible periodicities in the  IBIS/ISGRI data of the bright flare, which  eventually could  be 
interpreted as the possible pulse period of a neutron star compac object. A very fine  bin time (0.1 s) ISGRI  light  curve  (18--60 keV)  was
extracted from the flare  using the ii-light tool in OSA 10.2,  and  Solar system barycenter correction was applied to the photon arrival times. 
Periodicities were searched in the frequency range from 0.000628 Hz ($\sim$1590 s, after which the sensitivity is reduced due to the
finite length of the ISGRI light curve) to 5 Hz (0.2 s, corresponding to the Nyquist frequency of the data set).  Power spectra were  generated 
using the  fast Fourier transform analysis and the Lomb-Scargle periodogram technique  
but no  statistically significant evidence for coherent modulation was found. We estimated a 3$\sigma$ upper limit 
to the pulsed fraction of $\sim$ 5$\%$ (18--60 keV).
Given  the exceptionally high  flux of the source, we have also performed a timing analysis without the image deconvolution, in a non binning way (i.e. starting from the single events). This method is particularly suited to search  for very fast pulsations, e.g. up to few milliseconds. 
To optimize the search, we selected events  according to the Pixel Illuminated Factor (PIF) which is the 
fractional area of each pixel exposed to the source.  
In particular, only photons from pixels fully illuminated by the source (PIF=1)  were considered. 
By applying such PIF filter, it is possible to reduce the background and so to increase the signal to noise ratio. 
After barycentric correction of the photon arrival times in the original event lists, 
we performed both  Lomb-Scargle and fast Fourier transform analysis searching for periodicities as above, however no
significant evidence for a peak in the power spectra was found. 

\begin{figure}
\begin{center}
\includegraphics[height=6.5cm]{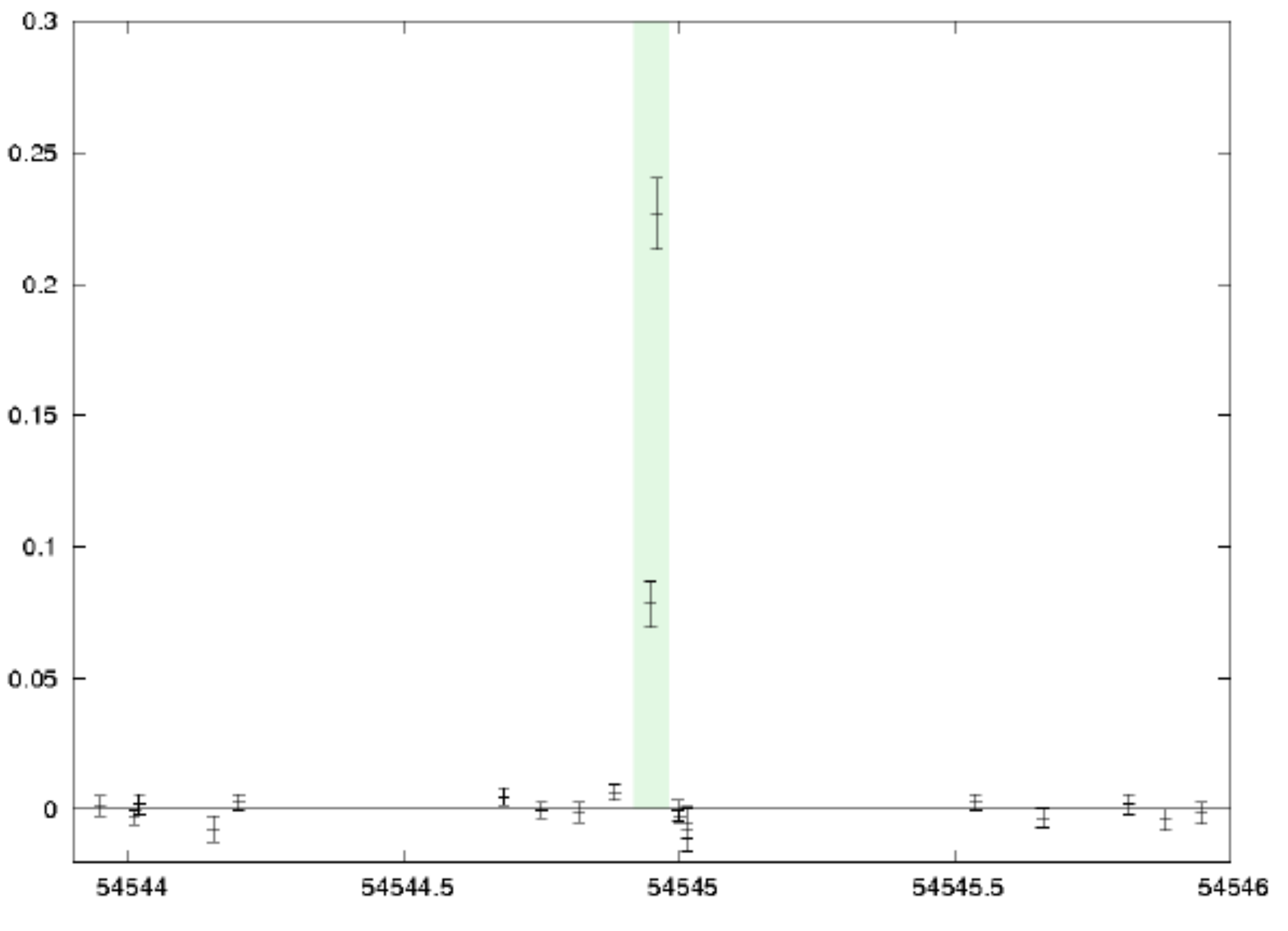}
\caption{Swift/BAT light curve  (15--50 keV)  on orbital  time-scale of the flare from IGR~J16479$-$4514. 
The shadowed green area indicates the portion of the flaring activity covered by the IBIS/ISGRI observation during the single ScW n. 43.
Time on the X-axis is in MJD unit.}
\end{center}
\end{figure}

\subsubsection{Targeted observation}

{\itshape INTEGRAL } performed a targeted observation of  IGR~J16479$-$4514 during revolution 1203,  from  2012 August  21 04:30:00 UTC to  21 August 21:25:42 UTC.  The source was in the fully coded FoV of IBIS/ISGRI for a total  effective on-source exposure of $\sim$ 16 ks.
As shown in Fig. 8 (where the orbital phases corresponding to the start and end time of the {\itshape INTEGRAL} observation are indicated by means of two blue lines)  almost the entire observation  took place  during the eclipse, when the compact object is  totally eclipsed by the supergiant companion donor.  

The IBIS/ISGRI mosaic was made in the energy band  22--60 keV (to take into account the evolution of the IBIS/ISGRI energy threshold that occurred from revolution number 900 on).  The source was not detected in the  mosaic, nor in any single ScW. As a result, we inferred   a 22--60 keV 3$\sigma$ upper limits  of 4.6 mCrab ($\sim$5$\times$10$^{-11}$ erg cm $^{-2}$ s$^{-1}$).  For comparison, we note that the lowest out-of-outburst as well as  out-of-eclipse hard X-ray emission   detected by IBIS/ISGRI  is known to be of the order of $\sim$1.7$\times$10$^{-11}$ erg cm $^{-2}$ s$^{-1}$ (20--60 keV, Sguera et al. 2008, Walter \& Zurita 2007), i.e. only a factor of 3 lower than this upper limit during the eclipse.  
 
\begin{figure}
\begin{center}
\includegraphics[height=4.5cm]{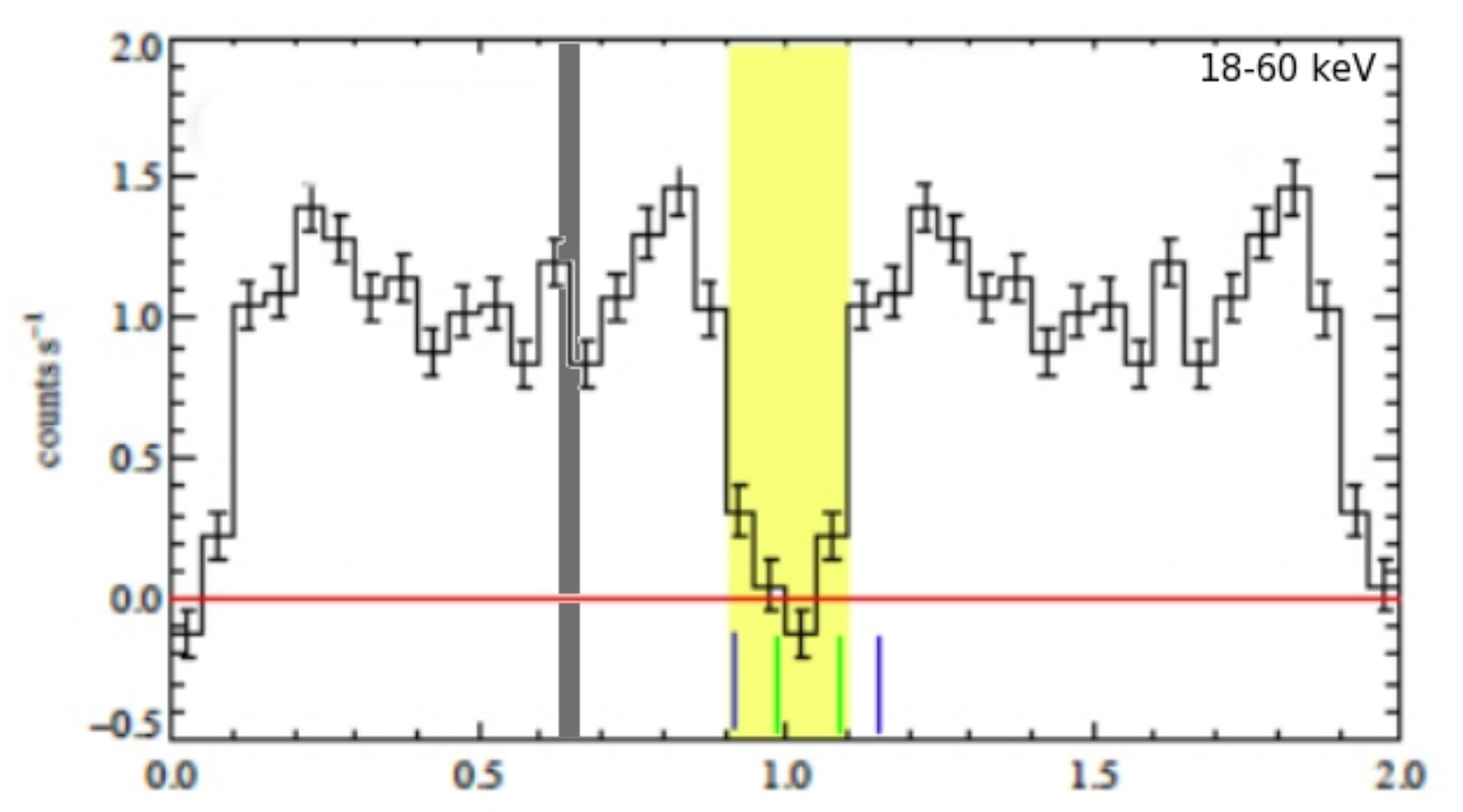}
\caption{IBIS/ISGRI light curve of IGR J16479$-$4514 (18--60 keV) phase-folded on the orbital period. 
Phase 0 is defined at the time of mid-eclipse MJD 55081.57 according to the most recent and precise ephemerides of Coley et al. (2015).  The shadowed yellow area indicates the entire duration of the X-ray eclipse. 
The grey thick line marks the occurrence of the powerful flare detected by IBIS/ISGRI. The two blue lines (green lines) indicate the start time and the end time of the targeted {\itshape INTEGRAL} ({\itshape XMM-Newton}) observation.}
\end{center}
\end{figure} 

\section{Discussion}

\subsection{X-ray eclipse} 

During the {\itshape XMM-Newton} observation performed in 2012, no point-like X-ray emission was detected from IGR~J16479$-$4514. However, extended X-ray emission was clearly observed up to a distance of at least 40 arcsec. As already noted by Bozzo et al. (2008), such a size is compatible with a dust scattering halo produced by the X-ray emission of IGR J16479$-$4514 before being eclipsed by its companion. Also its spectrum, significantly softer than the typical spectrum of this X-ray source out of eclipse, is consistent with that expected from dust-scattering, whose efficiency steeply decreases with photon energy  (see, e.g., Figure 6 in Draine 2003). 

From a time-resolved spectral analysis of the 2008 observation, performed immediately after a bright flare and during earlier phases of the eclipse, we could clearly disentangle the scattering halo component from two distinct components from the central point source: the brightest one is detected only during the initial phases of the eclipse and is characterized by a very significant excess of photoelectric absorption and can therefore be attributed to the direct emission from IGR J16479$-$4514 passing through a dense local environment. The second point-like component, instead, is detected throughout the full 2008 observation and declines faster than the dust scattering halo (see Fig.~\ref{fig:fluxes}); its spectral slope is comparable to the direct component, but its absorption is much smaller and, considering also the large equivalent width of the detected emission lines, can be attributed to scattering in the stellar wind. The largest absorption of the direct component suggests the presence of an absorbing structure in the orbital plane, which does not obscure the scattered emission, emerging more clearly during eclipse.  A spatially larger but much less dense structure surrounding IGR~J16479$-$4514 might be responsible for part of the absorption affecting all the spectral components, since its value of \nh=(6$\pm$2)$\times$10$^{22}$ cm$^{-2}$ is significantly larger than the total Galactic absorption column expected in this sky direction (\nh$\sim$2$\times$10$^{22}$ cm$^{-2}$; Willingale et al. 2013). However, we note that a similar excess is measured in two nearby bright X-ray binaries, GX~340$+$0 (Miller et al. 2016) and XTE~J1652--453 (Hiemstra et al. 2011) and therefore its origin might be the complex structure of the interstellar medium in this Galactic region rather than the local environment of IGR~J16479$-$4514.
 
The comparison of the fluxes of the different components shows how the contribution from the dust scattering halo increases at later eclipse phases (see Fig.~\ref{fig:fluxes}). Moreover, the flux of the halo during the 2012 observation is more than 3 times smaller than during the 2008 observation, which indicates that also the flux of IGR J16479$-$4514 out of eclipse was significantly smaller in the hours/days preceding the most recent observation. For these reasons, in 2012 the dust scattering halo could be observed without apparent contamination from the central point source.

\subsection{Giant hard X-ray flare} 
 {\itshape INTEGRAL}  detected a fast hard X-ray flare from IGR~J16479$-$4514 ($\sim$25 minutes duration) which reached a remarkable peak flux of  
2.10$\pm$0.26 Crab (18--60 keV). It translates into a X-ray luminosity of (7.0$\pm$0.9)$\times$10$^{37}$ erg  s$^{-1}$  if we assume the most constrained value of 4.6 kpc for the distance, to date (Coley et al. 2015).  This is  the most powerful flare ever detected from the source. Moreover, it is  one of the most powerful flares ever detected from any SFXT, in this respect  being second  only to the giant soft X-ray flare  detected by {\itshape Swift/XRT} from the SFXT  IGR~J17544$-$2619  (Romano et al. 2015), which reached a peak flux (luminosity) of  2.1 Crab (3$\times$10$^{38}$ erg  s$^{-1}$)  but in the softer  energy band 0.3--10 keV, so a proper comparison between the two  events is not possible. Its physical origin was explained in the framework of accretion from a transient disk (Romano et al. 2015).  Previously to our  current work, the dynamic range of the source  above 20 keV was  of the order of $\sim$350, as reported in the literature (Sguera et al.2008) by considering the highest measured luminosity level in outburst (1.5$\times$10$^{37}$ erg  s$^{-1}$ at 4.6 kpc distance, 20-60 keV)  and the lowest detected hard X-ray emission level outside outburst (4.3$\times$10$^{34}$ erg  s$^{-1}$ at 4.6 kpc distance, 20-60 keV). Our reported giant hard X-ray flare, the strongest ever, allows us to  push further the dynamic range up to a value of $\sim$1630.

For the sake of completeness, we note that this  flare from IGR~J16479$-$4514 has been previously studied with {\itshape Swift} by Romano et al. (2008). The flare triggered the BAT monitor on board {\itshape Swift} on 2008 March 19 at 22:44:45 UT (Barthelmy et al. 2008). Subsequently the satellite slewed to the target  with the XRT instrument, however its  observation temporally covered  the flare only during its rising  for  about 1,000 seconds (see shadowed green area in Fig. 5). Romano et al. (2008) mainly performed a   {\itshape Swift}  broad band X-ray spectral study of this flare. Here we note that, because of the  partial \sw~temporal coverage of the flare which was limited only to its  rising part, it was not possible to provide any exact measurement  on its  duration, peak-time and peak-flux. Conversely, our newly reported  \inte~detection fully covered the entire duration of the flare allowing such measurements and unveiling  its very energetic nature. In this respect our reported {\itshape INTEGRAL}  results are novel and complementary with  those previously reported from the \sw~observation (Romano et al. 2008). 

This exceptionally energetic flare raises  questions on its physical origin. Within the quasi spherical settling accretion model of Shakura et al. (2012, 2013, 2014), the production of bright X-ray flares from SFXTs (i.e. L$_x>10^{36}$ erg  s$^{-1}$) is triggered   by sporadic capture of magnetized stellar wind plasma by the magnetized compact object. The consequent magnetic reconnection increases the magnetospheric plasma entry rate, this results in abundant production of X-ray photons, strong Compton cooling and ultimately in the unstable accretion of the entire quasi-static shell previously formed above the neutron star magnetosphere. Clearly the shell can be reformed by new wind capture so the flares production can be repeated as long as the rapid mass entry rate into the magnetosphere is sustained.  In this scenario, a bright flare must be produced on the free fall time scale of the shell (i.e. $\sim$ 1,000--10,000 s) with  a typical energy released of $\sim$ 10$^{39 }$ ergs, corresponding to a typical mass of the shell  of $\sim$ 10$^{19}$ g. This view is broadly consistent with the typically measured energy released in known SFXTs bright flares detected by {\itshape INTEGRAL} (in the range 10$^{38}$--10$^{40}$ ergs) as well as with the corresponding measured  mass fallen onto the compact object (10$^{18}$--10$^{20}$ g) (Shakura et al. 2014). As for the specific case of IGR~J16479$-$4514,  we took into account  the peak luminosity of the giant hard X-ray flare and  accordingly calculated the corresponding  released energy and  accreted mass  following  Shakura et al. (2014). We derived values of $\sim$5$\times$10$^{41}$ ergs and $\sim$6$\times$10$^{21}$ g, respectively.  Clearly, this is an exceptional event with respect to the  usual flares typically detected, since it released a much greater amount of energy and  it required a much larger  amount of accreted material than usual. 

IGR~J16479$-$4514 is known to undergo a regular flaring activity  at a specific orbital phase $\phi$$\sim$0.65 (if the  epoch MJD 55081.57 is assumed as $\phi$=0 according to Coley et al. 2015). This  interesting behaviour  has been  firstly reported by Bozzo et al. (2009) using {\itshape Swift} observations and subsequently confirmed  by Sidoli et al. (2013)  through a nearly complete orbital monitoring  with {\itshape Suzaku}. This characteristic likely indicates the presence of a phase-locked large-scale structure in the supergiant wind, consecutively the flares are likely triggered by higher accretion rate onto the compact object during its passage inside of it. We note that such large-scale structures are believed to be an ubiquitous characteristic of all isolated supergiant stars (Puls et al. 2008, Massa \& Prinja 2015). In particular  Corotating Interaction Regions (CIRs) have been suggested as a potential candidate (Mullan 1984, Cranmer \& Owocki 1996), being characterized by an over-density with respect to  the  unperturbed surrounding stellar wind. 
 Since the giant flare detected by \inte~occurred at orbital phase $\phi$=0.648$\pm$0.002 (if the  epoch MJD 55081.571 is assumed as $\phi$=0
and 3.31961$\pm$0.00004 days is assumed as orbital period according to the most recent and  precise results of Coley et al. 2015), we propose  that  the interaction of the compact object with a CIR in the supergiant wind, during its passage inside of it,  has provided the needed very large amount of accreted material.

In order to support our hypothesis, we note that   Corbet \& Krimm  (2013) reported the light curve of 
IGR~J16479$-$4514 folded on its superorbital period of $\sim$11.88 days (see their figure 8), it clearly shows a sharp peak at superorbital phase in the range $\Phi$=0--0.08 (if phase $\Phi$=0 is assumed at  MJD 55996 according to Corbet \& Krimm  2013). Recently  Bozzo et al. (2017) proposed that the interaction between the compact object  and the CIRs in the wind of its supergiant  companion could drive the  superorbital periodicities observed in some SFXTs as well as SGXBs.  When the NS encounters the CIR, the different
velocity and density of this structure compared
to the surrounding stellar wind produces the required
long-term variation of the mass accretion rate to give
rise to a super-orbital modulation with the observed intensity. Under this hypothesis,  we calculated for IGR~J16479$-$4514 a CIR rotating period of $\sim$4.607 days according to equation 1 in Bozzo et al. (2017), with the assumption  that there is a single arm CIR  in the stellar wind. However, we note that it cannot be excluded apriori the possibility of multiple n CIRs  (in this case the rotation period would be 4.607/n). We used the value of 4.607 days  to calculate the superorbital phase corresponding to the  giant hard X-ray flare detected by {\itshape INTEGRAL}, and  we found a value of $\Phi$=0.035$\pm$0.021. Remarkably, we note that this is in agreement with the superorbital phase of the peak (in the range $\Phi$=0--0.08) corresponding to maximum flux  in the light curve of IGR~J16479$-$4514 folded on its superorbital period (see figure 8 in  Corbet \& Krimm  2013). This supports our hypothesis  that the giant hard X-ray flare has been produced likely during  the passage of the compact object inside the putative CIR. As a further check, in Fig. 9 we shows the IBIS/ISGRI light curve (18--60 keV) folded on the superorbital period according to  Corbet \& Krimm  (2013). The used light curve covers a time range of $\sim$13 years of {\itshape INTEGRAL} observations, from January 2003 to January 2015. We note that the shape of the folded {\itshape INTEGRAL} light curve  is broadly consistent with that of {\itshape Swift/BAT} as published by Corbet \& Krimm  2013 (see their Fig. 8) in a similar energy range (15--50 keV), i.e. it is characterized by a relatively sharp rise from minimum to maximum followed by a less intensive plateau phase. However, in addition we point out that the shape of the folded {\itshape INTEGRAL} light curve is bumpier than that of {\itshape Swift/BAT} since it shows some extra peaks which are very likely due to hard X-ray  flares from the source presumably detected by {\itshape INTEGRAL}, which on the contrary were not detected by {\itshape Swift/BAT} due to its  higher instantaneous sensitivity, with respect to {\itshape INTEGRAL},  on the short time scale typical of the detected flares (i.e. 2,000 seconds).

We note  that the putative  arm CIR  should likely be a structure stable over  timescales of several  years, as suggested by the detection with {\itshape INTEGRAL}, {\itshape Swift/BAT} and {\itshape Suzaku} of several X-ray flares at $\phi$  $\sim$ 0.65 spanning a  temporal range of at least $\sim$ 7 years.  The stability is further corroborated by the persistent detection of the  superorbital modulation in the {\itshape Swift/BAT}  light curve covering a timescale of about 8.5 years (Corbet et al. 2013). 
The giant flare detected by {\itshape INTEGRAL} is exceptionally energetic compared to typical flares observed at the same orbital phase, this could imply changes of the mass accretion rate related to cycle-to-cycle variations of the arm CIR density (e.g. as in  the case of the wind fed supergiant HMXB IGR J16493$-$4348, Coley et al. 2019).

\begin{figure}
\begin{center}
\includegraphics[height=5.3cm]{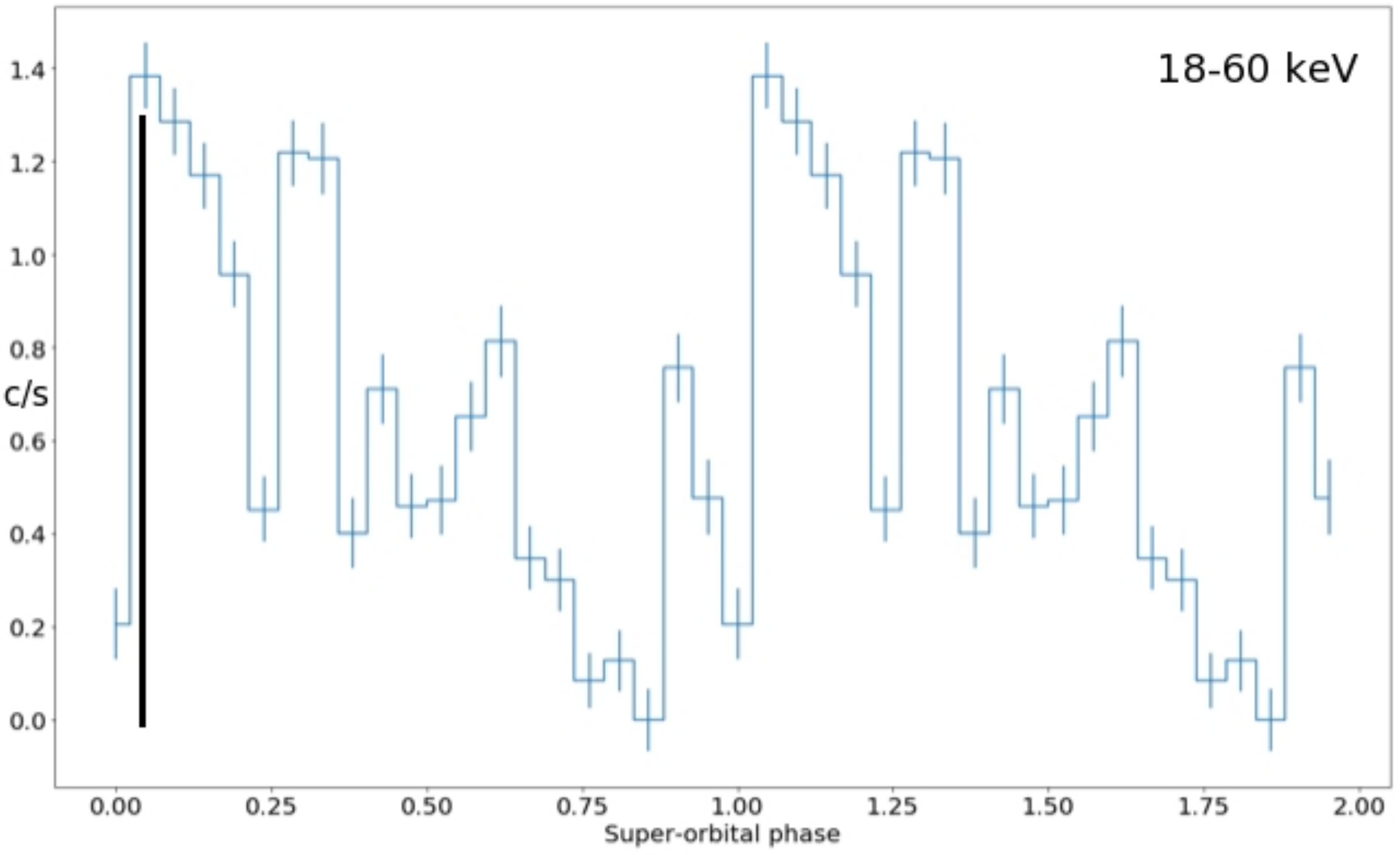}
\caption{IBIS/ISGRI light curve of IGR J16479$-$4514 (18--60 keV) phase-folded on the superorbital period. 
Phase 0 is defined according to Corbet \& Krimm  2013.
The black  line marks the occurrence of the powerful flare detected by  {\itshape INTEGRAL}.}
\end{center}
\end{figure}

\section{Summary and conclusions}
In this work we presented new {\itshape XMM-Newton} and {\itshape INTEGRAL} results on the SFXT 
IGR J16479$-$4514. The main results and conclusions from our analysis are the following:

(i) no point source was detected during the unpublished 2012  {\itshape XMM-Newton} observation, performed during eclipse. This allowed us to clearly detect, 
without any contamination, the diffuse soft X-ray emission discovered
in the  previously published  2008 {\itshape XMM-Newton} observation during eclipse
and tentatively interpreted as a dust-scattering halo (Bozzo et al. 2008). The significantly lower flux observed in the 2012 observation strongly supports the dust halo interpretation.

(ii) a time-resolved re-analysis of the archival 2008 {\itshape XMM-Newton} observation,  adopting more time intervals and source extraction regions (optimized either for point-like and diffuse emission) than in Bozzo et al. (2008), has allowed us to better characterize the point-like emission during eclipse, disentangling two components with different evolution timescales and power-law spectra with the same slope but significantly different photoelectric absorption.

(iii) {\itshape INTEGRAL}  detected an exceptionally energetic  (peak luminosity of $\sim$7$\times$10$^{37}$ erg  s$^{-1}$, 18--60 keV) and fast (25 minutes duration)  giant hard X-ray flare. They are very rare from SFXTs, this would be only the second one reported in the literature after that from  the SFXT  IGR~J17544$-$2619  detected in a much  softer energy band (peak luminosity of $\sim$2$\times$10$^{38}$ erg  s$^{-1}$, 0.2--10 keV)

(iv)  the detected giant hard X-ray flare required an  amount of accreted material ($\sim$7$\times$10$^{21}$ g) about two orders of magnitude greater than usually measured from all previous detected flares from the source. The giant flare occurred at a specific  orbital phase  
which has been previously suggested  to be linked to the presence of a stable large-scale structure in the supergiant wind. 
We invoke CIRs in the wind as a potential candidate  and consecutively we propose that the interaction of the compact object with such CIRs,  during its passage inside of it, have provided the needed very large amount of accreted material necessary  to produce the exceptionally energetic hard X-ray flare.

\section*{Acknowledgments}
We would like to thank the referee for valuable  comments which helped us to improve the manuscript. 
LS thanks M. Marelli for interesting discussions. AT acknowledges funding in the framework of the project ULTraS ASI--INAF contract N.\,2017-14-H.0.


\bibliographystyle{aasjournal}

\end{document}